\documentclass[showpacs, pra,twocolumn,preprintnumbers ,amsmath, amssymb, superscriptaddress, aps]{revtex4-2}
\usepackage{color}
\usepackage{amsmath,amssymb}
\usepackage{pifont}
\usepackage{amssymb}  
\usepackage{bbold}
\usepackage{float}
\usepackage{subfloat}
\usepackage{adjustbox}

\usepackage[caption=false]{subfig}
\usepackage{tikz}
\usepackage{makecell}
\usepackage{subfig}
\usepackage{pifont}   
\usepackage{graphicx} 
\graphicspath{{Figures/}}
\usepackage{dcolumn}  
\usepackage{bm}       
\usepackage{multirow} 
\usepackage{placeins}
\usepackage[colorlinks]{hyperref}
\usepackage{mathtools}
\usepackage{appendix}

\def \be{\begin{align}}
	\def \ee{\end{align}}
\def \bea{\begin{eqnarray}}
	\def \eea{\end{eqnarray}}

\begin{document}	
	\title{Tunable Goos–Hänchen shifts and group delay time in single-barrier silicene}
	\date{\today}
	
	\author{Youssef Fattasse}
	\email{a.jellal@ucd.ac.ma}
	\affiliation{Laboratory of Theoretical Physics, Faculty of Sciences, Choua\"ib Doukkali University, PO Box 20, 24000 El Jadida, Morocco}
	\author{Hocine Bahlouli}
	\affiliation{Physics Department and IRC Advanced Materials$,$
		King Fahd University
		of Petroleum $\&$ Minerals$,$
		Dhahran 31261$,$ Saudi Arabia}
	\author{Clarence Cortes}
	\affiliation{Vicerrector\'ia de Investigaci\'on y Postgrado, Universidad de La Serena, La Serena 1700000, Chile}  
	\author{David Laroze}
	\affiliation{Instituto de Alta Investigación, Universidad de Tarapacá, Casilla 7D, Arica, Chile}
	
	\author{Ahmed Jellal}
	\email{a.jellal@ucd.ac.ma}
	\affiliation{Laboratory of Theoretical Physics, Faculty of Sciences, Choua\"ib Doukkali University, PO Box 20, 24000 El Jadida, Morocco}

	\pacs{ 73.22.Pr, 72.80.Vp, 73.63.-b\\
		{\sc Keywords}: Silicene, rectangular barrier, Dirac equation,
		transmission, Goos–Hänchen shifts, group delay time.
	}
	
	\begin{abstract}
		
		We investigate the Goos–Hänchen (GH) shifts and group delay time of Dirac fermions traversing a rectangular electrostatic potential barrier in silicene. By analyzing their dependence on the incident angle, barrier height, barrier width, and incident energy, we demonstrate that the GH shifts exhibit pronounced oscillations arising from quantum interference within the barrier region. The amplitude and number of oscillation peaks increase with increasing energy, barrier width, and incidence angle, resulting in enhanced lateral beam displacement. Meanwhile, the group delay time exhibits resonant features associated with the formation of quasi-bound states, increasing with barrier width, energy, and incidence angle, while decreasing with increasing barrier height. These results clarify how barrier-induced quantum interference controls both the lateral and temporal dynamics of Dirac fermions in silicene, highlighting the potential role of electrostatic barriers in enabling tunable transport in two-dimensional Dirac materials.

	\end{abstract}

	\maketitle
	
	
	\section{Introduction}
	
	
	The first isolation of graphene in 2004 \cite{1} marked the beginning of intensive research into two-dimensional materials, which have since become central to both fundamental studies and technological applications. Researchers have been actively engaged in discovering new atomically thin materials that exhibit properties comparable to or complementary with the exceptional electronic, mechanical, and optical characteristics of graphene. Since then, the field expanded rapidly to cover various domains, which include advanced transistors \cite{Schwierz,Das},  topological field-effect transistors (FETs) \cite{Liu,Qian} and innovative device designs while it drives progress in condensed matter physics \cite{Castro} and materials chemistry \cite{Yuan,Kalantar}. Group-IV materials which resemble graphene have gained interest because they offer potential to work with current semiconductor manufacturing methods.
	
	Silicene is considered one of the most promising materials in this class \cite{Kara,Xu}. It exhibits a honeycomb lattice structure similar to graphene and a Dirac-like electronic dispersion; however, important differences arise from the intrinsic properties of silicon atoms. In particular, silicene has a slightly buckled structure, which leads to relatively strong spin–orbit coupling and enables the band gap to be tuned by an external electric field. These characteristics make the material highly suitable for applications in nanoelectronics and spintronics.
		The existence of silicene was first predicted by Takeda and Shiraishi in 1994 \cite{Takeda}. it was later studied in greater detail by Guzmán-Verri and co-workers, who introduced the term “silicene” in 2007 \cite{Guzman}. Silicene does not exist in nature because it lacks both natural occurrence and stable bulk allotrope that corresponds to graphite, which distinguishes it from graphene. The process of synthesizing and stabilizing this material becomes extremely difficult because traditional exfoliation techniques cannot be used.

	The lattice structure of silicene differs fundamentally from that of graphene, as it adopts a buckled configuration rather than a perfectly planar one. The two sublattices show a vertical displacement of about \(d \approx 0.46~\text{\AA}\) \cite{6,7} due to the fact that silicon possesses a larger atomic radius. The resulting buckling produces a complex hybridization pattern that combines both \(sp^{2}\) and \(sp^{3}\) hybridization orbitals. In contrast, graphene exhibits only \(sp^{2}\) hybridization bonding, ensuring its flat structure \cite{Pulci}. Silicene exhibits an electrically tunable band gap of approximately 1.55 meV \cite{Feng}, while its intrinsic spin--orbit coupling strength reaches 3.9 meV \cite{Jiang}. These properties make silicene a highly promising candidate for applications in both spintronic and valleytronic technologies.
	The buckled structure further enables efficient control of the electronic properties through an external perpendicular electric field, which induces a staggered on-site potential between the two sublattices \cite{6,7}. By tuning this field, silicene can undergo a topological phase transition from a quantum spin Hall insulating phase to a trivial band insulator. Such electrically driven topological transitions give rise to rich physical phenomena and offer promising routes towards robust information processing and quantum device applications \cite{Tahir,Kim}.

	The present work builds upon our previous studies of Goos–Hänchen (GH) shifts and group delay time in graphene under various external constraints \cite{Jel1,Jel2,Jel3}, and extends our more recent investigation of the tunneling behavior of fermions in silicene through electrostatic potential barriers \cite{JellalAP2024}. While our earlier silicene study focused primarily on how electrostatic barriers influence transmission and tunneling probabilities, the present analysis moves beyond stationary transport characteristics to explore the spatial and temporal dynamics of Dirac-fermion wave packets. In particular, we examine the GH shifts and group delay time that arise when Dirac fermions propagate through a rectangular potential barrier, and analyze their dependence on key system parameters such as the angle of incidence, barrier height, barrier width, and incident energy. This extension provides a more comprehensive understanding of barrier-induced transport in silicene, revealing how quantum interference effects, phase accumulation, and the formation of quasi-bound states within the barrier jointly govern both the lateral displacement of the transmitted beam and the associated time-delay phenomena. By linking tunneling behavior with spatial and temporal propagation characteristics, our results deepen the physical insight into wave-packet dynamics in buckled Dirac materials and highlight the role of electrostatic barrier engineering as an effective means of controlling transport in silicene-based nanoelectronic and valleytronic devices.
	
	The paper is organized as follows. In Sec.~\ref{MTM}, we introduce the theoretical model and present the basic formalism used to describe Dirac fermions in silicene in the presence of a rectangular electrostatic potential barrier. Section~\ref{TPB} is devoted to the analysis of the transmission properties, where the reflection and transmission coefficients are derived and discussed in detail. In Sec.~\ref{GHGD}, we focus on the central results of this work, namely the Goos–Hänchen shifts and the group delay time, and examine their dependence on the relevant system parameters. In Sec.~\ref{NAA}, we present and comment on the numerical results for various choices of the physical parameters that specify our model. We provide a comparison of the silicene and graphene results in Sec.~\ref{SSGG}. Finally, we conclude with a summary of the physical and applied aspects of our results.

	\section{Theoretical model}\label{MTM}
	
\begin{figure}[ht!]
		\centering
		\includegraphics[scale=0.16]{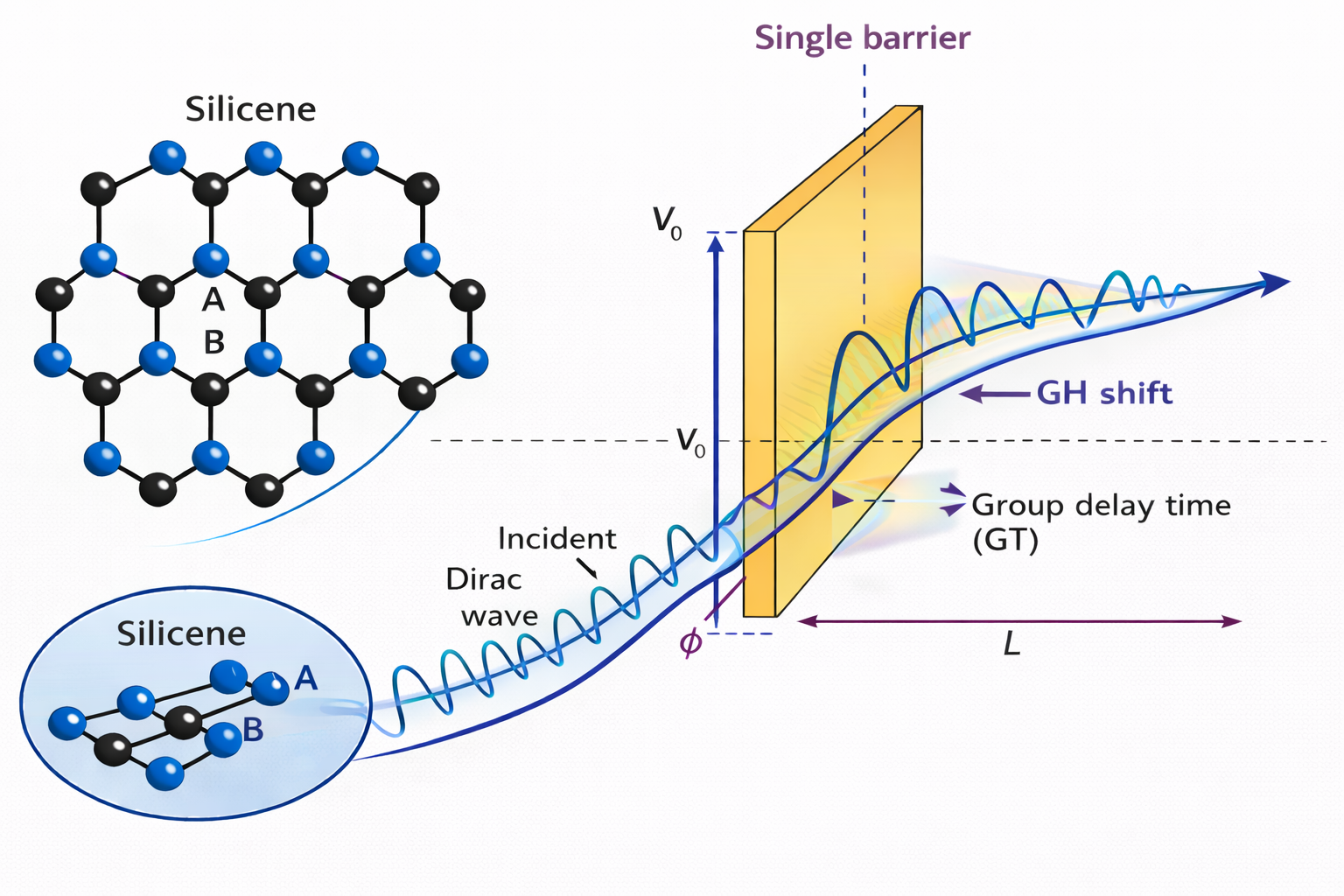}
		\caption{Schematic illustration of the Goos–Hänchen (GH) shift and group delay time for Dirac fermions traversing a single electrostatic potential barrier in silicene. The upper-left panel shows a top view of the silicene honeycomb lattice with two inequivalent sublattices \(A\) and \(B\), while the lower-left inset presents a side view highlighting the intrinsic buckled structure of silicene. On the right, an incident Dirac wave packet impinges on a rectangular electrostatic potential barrier of height \(V_0\) and width $L$ at an angle \(\phi\). Upon transmission, the wave packet undergoes a lateral displacement along the interface (GH shift) and experiences a finite group delay time due to multiple reflections and phase accumulation inside the barrier.}
		\label{fig1}
	\end{figure}

	Figure~\ref{fig1} illustrates the physical origin of the Goos--Hänchen shift and group delay time in silicene. An incident Dirac fermion, modeled as a wave packet, approaches a single electrostatic barrier at an oblique angle $\phi$. Quantum interference due to multiple reflections inside the barrier region take place and lead to phase accumulation that produces a lateral displacement of the transmitted beam along the interface, known as the Goos-Hänchen shift. At the same time, the carrier experiences a finite residence time within the barrier, which manifests as a group delay time. From a theoretical point of view, the honeycomb lattice of silicene, with its two-sublattice structure, gives rise to Dirac-like dispersion and governs the phase-dependent transport processes that underlie both these spatial and temporal effects.
	
	The system under consideration uses an effective Hamiltonian which describes its low-energy excitations near the Dirac points located at the corners of the first hexagonal Brillouin zone with $\eta=\pm 1$ at points $K$ and $K'$ \cite{JellalAP2024}. It is given by
	\begin{equation}
		H= v_{F}\left( \sigma _{x}p_{x}-\eta \sigma _{y}p_{y}\right) -\eta \tau
		_{z}\triangle _{so}\sigma _{z}+V\left( x\right) \mathbb{I}_{2}  \label{1}
	\end{equation}
	where $v_F= 5.3 \times 10^5$ m/s is the Fermi velocity, $\sigma_i(i=x,y,z)$ are the Pauli matrices,  operating on the pseudospin (sublattice) space, and ($p_x,p_y$) are two components of the momentum operator  $\vec{p}$. The second term in~\eqref{1} represents the intrinsic spin--orbit coupling of the Kane--Mele type \cite{5}, which serves as a unique characteristic for silicene. The associated spin--orbit gap is $\Delta_{so}=3.9$~meV, and the index $\tau_z=\pm 1$ labels the two spin projections, spin-up ($\tau_z=+$) and spin-down ($\tau_z=-$). The last term describes the effect of an external electrostatic potential $V(x)$, which models a rectangular potential barrier of width $L$ and height $V_0$ as illustrated in Fig.~\ref{fig1}, and $\mathbb{I}_2$ is the $2\times 2$ unit matrix.The potential profile is defined by 
	\begin{equation}
		V\left( x\right) =V_{j}=\left\{ 
		\begin{array}{ccc}
			0,\text{ \ } & x<0 \\ 
			V_{0}, \text{ \ }& 0<x<L \\ 
			0,\text{ \ } & x>L.
		\end{array}%
		\right.   \label{2}
	\end{equation}
	In principle, the full low-energy Hamiltonian of silicene can be written in a 4×4 form that accounts for both spin and valley degrees of freedom. However, because the spin–orbit interaction appears as a diagonal mass-like term and the valleys are decoupled near the Dirac points, the Hamiltonian becomes block-diagonal. Therefore, for fixed valley index $\eta=\pm1$ and spin projection $\tau_z=\pm1$, the problem reduces to an effective 2×2 Dirac Hamiltonian. This reduction is widely employed in transport studies of silicene and allows 
		each spin-valley sector to be  treated independently, see for instance \cite{JellalAP2024} and references therein.

	The commutation relation $[H, p_y] =0$ allows us to separate out the $y$-component of the spinor wave function, which leads to the representation of eigenspinors through plane waves that travel in the $y$-direction according to the formula $\psi_{j}(x,y)=e^{ik_{y}y}(\phi _{j}^{+}(x),\phi _{j}^{-}(x))^{T}$.  The superscript ${T}$ represents the transpose of a row vector.
	We now solve the eigenvalue equation \(H\psi_{j}=E\psi_{j}\) in each region $j=1,2,3$ for a given valley index \(\eta\) and spin projection \(\tau_{z}\). In region~1 (\(x<0\)), where the electrostatic potential vanishes, the eigenspinors are written as a superposition of incident and reflected wave components
	\begin{widetext}
		\begin{equation}
			\psi _{1}(x,y)=\frac{e^{ik_{y}y}}{\sqrt{DW\left( 1+\alpha ^{2}\right) }}%
			\left[  
			\begin{pmatrix}
				1 \\ 
				s_{1}\alpha e^{-i\eta \phi }%
			\end{pmatrix} e^{ik_{x}x}+r
			\begin{pmatrix}
				1 \\ 
				-s_{1}\alpha e^{i\eta \phi }%
			\end{pmatrix} e^{-ik_{x}x}\right]  \label{4}
		\end{equation}
	\end{widetext} 
	where the parameters \(D\) and \(W\) define the width and length measurements of the silicene sheet in \(x\)- and \(y\)-dimensions. The coefficient \(r\) denotes the reflection amplitude,  the parameter $\alpha$ and icident angle $\phi$ are
	\begin{align}
		&\alpha =\sqrt{\frac{E_{1}+\eta \tau _{z}\triangle _{so}}{E_{1}-\eta \tau
				_{z}\triangle _{so}}}\\
		&\phi =\arctan \left(\frac{k_{y}}{k_{x}}  \right)       
	\end{align}
	with the sign function $s_{1}=\text{sgn}(E_{1}-\eta \tau
	_{z}\triangle _{so})$ refers to the conduction and valence bands.
	The longitudinal and transverse components of the particle wave vector are readily obtained as
	\begin{equation}
		k_x = k \cos\phi, \quad k_y = k \sin\phi 
		\label{6}
	\end{equation}
	which lead to the dispersion relation
	\begin{equation}
		E_1 = s_1 \hbar v_F \sqrt{k^2 + \Delta_{so}^{\,2}} .
		\label{7}
	\end{equation}
	Here, \(k = |\vec{k}|\) denotes the magnitude of the wave vector, with \(\vec{k}=(k_x,k_y)\).
	We show that the eigenspinor
	in region $2$ takes the form
	\begin{widetext}
		\begin{equation}
			\psi _{2}(x,y)=\frac{e^{ik_{y}y}}{\sqrt{DW\left( 1+\beta ^{2}\right) }}\left[
			c
			\begin{pmatrix}
				1 \\ 
				s_{2}\beta e^{-i\eta \theta }%
			\end{pmatrix} e^{iq_{x}x}+d
			\begin{pmatrix}
				1 \\ 
				-s_{2}\beta e^{i\eta \theta }%
			\end{pmatrix} e^{-iq_{x}x}\right]  \label{8}
		\end{equation}%
	\end{widetext}
	where we have set
	\begin{align}
		&  \beta =\sqrt{\frac{E_{2}-V_{0}+\eta \tau _{z}\triangle _{so}}{%
				E_{2}-V_{0}-\eta \tau _{z}\triangle _{so}}} \\
		& \theta =\arctan \left(\frac{k_{y}}{q_{x}}  \right)  
	\end{align}
	with  $c$ and $d$ are two constants, $s_{2}=\text{sgn}(E_{2}-\eta \tau
	_{z}\triangle _{so}-V_{0})$.  
	The two components of the wave vector $\vec q=(q_x, k_y)$  are 
	\begin{align}
		q_{x}=q
		\cos \theta, \quad
		& k_{y}=q
		\sin \theta  \label{9}
	\end{align}
	such that   {$q=|\vec q|$} is
	\begin{align}
		q=\frac{\sqrt{(E_{2}-V_{0})^{2}-\Delta _{so}^{2}}}{\hbar v _{F}}
	\end{align}
	leading to the energy $E_2$. In region 3, we obtain the eigenspinor
	\begin{equation}
		\psi _{3}(x,y)=\frac{te^{ik_{y}y}}{\sqrt{DW\left( 1+\alpha ^{2}\right) }}%
		\begin{pmatrix}
			1   \\ 
			s_{1}\alpha e^{-i\eta \phi}
		\end{pmatrix} e^{ik_{x}x}  \label{10}
	\end{equation}%
	and $t$ denotes the transmission coefficient.
	
	Owing to the stationary nature of the system, energy conservation requires that the total energy of the Dirac fermions remains the same in all regions, namely \(E = E_{1} = E_{2} = E_{3}\). In addition, translational invariance along the \(y\)-direction ensures the conservation of the transverse wave vector \(k_y\). As a result, using~\eqref{6} and \eqref{9}, a direct relation is established between the incident angle \(\phi\) in region~1 and the propagation angle \(\theta\) inside the barrier (region~2), which governs the refraction and scattering properties of the fermions at the barrier interfaces. This is 
	\begin{equation}
		\sin \theta =\sqrt{\frac{E^{2}-\Delta _{so}^{2}}{(E-V_{0})^{2}-\Delta
				_{so}^{2}}}\sin \phi .  \label{111}
	\end{equation}%
	For a specific incident energy, we identify a maximum angle that allows electrons to continue transmitting through the barrier. The limiting case occurs at a refraction angle of \(\theta = \pi/2\), which causes the longitudinal component of the wave vector to disappear within the barrier. When the incident angle exceeds this critical point, the \(x\) component of the wave vector becomes entirely imaginary, which causes the propagating states to change into evanescent modes that exist inside the barrier. The corresponding wave functions exhibit exponential decay along the transport path, which results in decreased transmission. The critical angle that distinguishes between propagating and evanescent regimes is established by applying the condition \(\theta = \pi/2\) to~\eqref{111}. It is given by
	\begin{equation}
		\phi _{l}=\arcsin \sqrt{1+\frac{V_{0}^{2}-2V_{0}E}{E^{2}-\Delta _{so}^{2}}}.
		\label{12}
	\end{equation}%
	The above results will be used to compute the transmission probability and, more importantly, to evaluate the Goos–Hänchen shift and the associated group delay time. These quantities are directly extracted from the transmission amplitude and allow us to analyze how lateral beam displacement and temporal delay emerge from phase accumulation and quantum interference effects at the barrier.
	
	\section{Transmission }\label{TPB}
	

	In our model, the Hamiltonian of silicene resembles  the effective spin-$\frac{1}{2}$ Dirac Hamiltonian used in graphene, in contrast to the spin-1  Hamiltonian of the $\alpha-T_3$ lattice \cite{Emilia2017, Iurov2020}. The scattering problem is solved by applying appropriate boundary conditions at the interfaces $x=0$ and $x=L$. Silicene has a honeycomb lattice with two sublattices (A and B), similar to graphene. Expanding the tight-binding model  near the Dirac points $K$ and $K'$ in the Brillouin zone yields long-wavelength effective spin-$\frac{1}{2}$ Dirac Hamiltonian, highlighting the similarity to graphene. 
		To impose appropriate boundary conditions, we require the continuity of the probability current $J_x$ perpendicular to the interface. This ensures conservation of probability and implies continuity of the two-component spinor wave function across the boundary, analogous to graphene-based Dirac systems. Consequently, the eigenspinors~\eqref{4}, \eqref{8}, and \eqref{10} are continuously matched at $x=0$ and $x=L$, resulting in a system of linear equations for the reflection and transmission amplitudes. Solving this system analytically provides the transmission and reflection coefficients directly. They are given by
\begin{widetext}
	\begin{align}
		t &=\frac{\left( 1+e^{2\eta i\theta }\right) \left( 1+e^{2\eta i\phi
			}\right) e^{-iL\left( k_{x}-q_{x}\right) }}{1+e^{2\eta i\left( \theta +\phi
				\right) }+e^{2i\left( \eta \theta +q_{x}L\right) }+e^{2i\left( \eta \phi
				+q_{x}L\right) }+\frac{s_{1}s_{2}\left( \alpha ^{2}+\beta ^{2}\right) \left[
				e^{i\eta \left( \theta +\phi \right) }-e^{i\left( \eta \left( \theta +\phi
					\right) +2q_{x}L\right) }\right] }{\alpha \beta }}  \label{14} \\
		r &=\frac{e^{-i\eta \phi }\left( -1+e^{2iq_{x}L}\right) \left[ -\alpha
			^{2}e^{i\eta \theta }+s_{1}s_{2}\alpha \beta \left( e^{i\eta \phi }-e^{i\eta
				\left( 2\theta +\phi \right) }\right) +\beta ^{2}e^{i\eta \left( \theta
				+2\phi \right) }\right] }{s_{1}s_{2}\alpha \beta \left[ 1+e^{i\eta 2\left(
				\theta +\phi \right) }+e^{i2\left( \eta \theta +q_{x}L\right) }+e^{i2\left(
				q_{x}L+\eta \phi \right) }\right] +\left( \alpha ^{2}+\beta ^{2}\right) %
			\left[ e^{i\eta \left( \theta +\phi \right) }-e^{i\left( \eta \left( \theta
				+\phi \right) +2q_{x}L\right) }\right] }  \label{15}.
	\end{align}%
\end{widetext}
The transmission and reflection probabilities are determined through the assessment of the current density corresponding to Dirac fermions in silicene. This process confirms that the calculated probabilities maintain appropriate particle movement across the boundaries while fulfilling the requirements of current flow conservation. The formalism defines transmission and reflection probabilities through the corresponding scattering amplitudes, which are expressed as 
\begin{align}
	T  = |t|^{2}, \quad R = |r|^{2}.
\end{align}
The equations establish a physical interpretation for the coefficients \(t\) and \(r\), which serve as indicators of transmission and reflection probabilities. The explicit analytical expressions for these probabilities emerge after we perform direct algebraic manipulations to get
\begin{widetext}
	\begin{align}
		T &=\frac{\cos ^{2}\theta \cos ^{2}\phi }{\left[ \cos \theta \cos \phi \cos
			\left( q_{x}L\right) \right] ^{2}+\sin^{2} \left( q_{x}L\right) \left[ \frac{%
				\alpha ^{2}+\beta ^{2}}{2\alpha \beta }-s_{1}s_{2}\sin \theta \sin \phi %
			\right] ^{2}}  \label{16} \\
		R &=\frac{\frac{\alpha ^{2}+\beta ^{2}}{\alpha \beta }\left[ \frac{\alpha
				^{2}+\beta ^{2}}{4\alpha \beta }-s_{1}s_{2}\sin \theta \sin \phi \right]
			+\sin^{2} \theta -\cos^{2} \phi }{\left[ \frac{\cos \theta \cos \phi }{\tan
				\left( q_{x}L\right) }\right] ^{2}+\left[ \frac{\alpha ^{2}+\beta ^{2}}{%
				2\alpha \beta }-s_{1}s_{2}\sin \theta \sin \phi \right] ^{2}}.  \label{17}
	\end{align}
\end{widetext}


\section{GH shifts and group delay time}\label{GHGD}

	The Goos--H\"anchen (GH) shift was originally introduced in optics as the lateral displacement of a linearly polarized electromagnetic beam upon reflection at an interface~\cite{199,200,211}. A similar phenomenon occurs in electronic systems: when the phase of the transmission or reflection amplitude depends on the transverse wave vector, a quantum wave packet can experience a lateral displacement. In this work, the GH shift corresponds to the sideways displacement of a transmitted Dirac-fermion wave packet in silicene, obtained using the stationary-phase approximation applied to the transmission phase. Such electronic GH shifts have been extensively studied in graphene and other Dirac materials, providing a useful framework for describing wave-packet dynamics during tunneling processes.
	Based on previous studies in graphene~\cite{Jel1,Jel2,Jel3}, where the Goos--H\"anchen (GH) shifts and group delay time have been extensively analyzed, we investigate how these phenomena manifest in silicene. In particular, we aim to explore the lateral displacement of Dirac-fermion wave packets and the corresponding temporal response as they tunnel through electrostatic barriers, highlighting similarities and differences with graphene and other Dirac materials.

To proceed, let us express  the transmission (\ref{14}) and reflection (\ref{15}) coefficients as 
\begin{align}\label{Cnum}
	t=\sqrt{T} e^{i\varphi_{t}}, \quad
	r=\sqrt{R}e^{i\varphi_{r}}
\end{align}
where  $T$ and $R$ are given in (\ref{16}-\ref{17}), while 
the corresponding  phase shifts are
\begin{align}
	\varphi_{t}=\arctan\left(i\frac{t^{\ast}-t}{t+t^{\ast}}\right), \quad \varphi_{r}=\arctan\left(i\frac{r^{\ast}-r}{r+r^{\ast}}\right).	
\end{align}
In what follows, we discuss the theoretical framework used to investigate the group propagation time in both transmission and reflection. In particular, a spatiotemporal wave packet can be employed to model a finite-duration electron beam as a weighted superposition of plane-wave spinors. Following \cite{chen08}, the incident, reflected, and transmitted beam waves at the interfaces 
$ (x = 0, x = D) $ can be expressed as double Fourier integrals over the frequency $\omega$ and the transverse wave vector $k_y$. Representative expressions for these wave packets are given by
\begin{align}
	&\label{int1}
	\Phi_{\sf in}(x,y, t)=\iint f(k_y,\omega)\ \Psi_{\sf in} (x,y) \  e^{-i\omega
		t}\ dk_yd\omega\\
	&\label{int2}
	\Phi_{\sf re}(x,y,t)=\iint  rf(k_y,\omega)\ \Psi_{\sf re} (x,y) \ e^{-i \omega
		t}\ dk_yd\omega
	\\
	&\label{int3}
	\Phi_{\sf tr}(x,y,t)=\iint  tf(k_y,\omega) \ \Psi_{\sf tr} (x,y) \ e^{-i \omega
		t}\ dk_yd\omega
\end{align}
in which the spinors $ \Psi_{\sf in}, \Psi_{\sf re} $ and  $ \Psi_{\sf tr} $ are  provided in \eqref{4}
and \eqref{10}.   According to \cite{Beenakker}, the angular spectral distribution is considered to have a Gaussian form $f(k_y,\omega)=w_ye^{-w_{y}^2(k_y-\omega)^2}$ with the half beam width at the waist being $\omega_y$, and the wave frequency is $\omega=E/\hbar$. Injecting \eqref{Cnum} into (\ref{int2}-\ref{int3}) to determine the total phases of the reflected and transmitted wave functions at ($x = 0 $, $x = D $). This process yields to 
\begin{equation}
	\mathbf{\Phi}_{r}=\varphi_{r}+k_yy-\omega t, \quad \mathbf{\Phi}_{t}=\varphi_{t}+k_yy-\omega t.
\end{equation}
Using the stationary phase approximation \cite{Steinberg, Li1}, i.e.  $  \frac{\partial\mathbf{\Phi}_{\gamma}}{\partial\phi}=0 $ and $  \frac{  \partial\mathbf{\Phi}_{\gamma}}{\partial\omega}=0 $, we calculate the GH shifts $ S_{\gamma} $  and  group delay time $ \tau_{\gamma} $
\begin{align}
	\label{ghss}	S_{\gamma}&=- \frac{\partial \varphi_{\gamma}}{\partial
		k_{y}}\\	
	\tau_{\gamma}&= \tau^{\varphi_{\gamma}} +
	\tau^{s_{\gamma}}=\frac{\partial \varphi_{\gamma}}{\partial
		\omega}+\left(\frac{\partial k_y}{\partial
		\omega}\right)S_{\gamma}
\end{align}
where $\lambda=t, r$ stands for the transmission and reflection amplitudes, such that
$ \tau^{\varphi_{\gamma}}$ denotes the time derivative of
phase shifts, while  $\tau^{s_{\gamma}}$  denotes the contribution of  $S_{\gamma}$. Note that 
$\tau_{\gamma}$  can be thought of as the average of the group delay times of the two components because the wave function involves a two-component spinor. As a result, we have 
\begin{eqnarray}
	\tau^{\varphi_{t}}=\hbar \frac{\partial \varphi_t}{\partial
		E}+\frac{\hbar}{2}\frac{\partial \phi'}{\partial E}, \quad
	\tau^{\varphi_{r}}=\hbar \frac{\partial \varphi_r}{\partial E}
\end{eqnarray}
and in GH shifts
\begin{equation}
	\tau^{s_{t}}=\frac{\sin\phi}{\upsilon_F}S_t, \quad
	\tau^{s_{r}}=\frac{\sin\phi}{\upsilon_F}S_r.
\end{equation}
Next, we will numerically  present and analyze the GH shifts $S_t$ and group delay time $\tau_{t}/\tau_0$ for Dirac fermions in silicene when
scattered by a potential barrier based on the results obtained previously. For this purpose, we introduce the scaled Fermi wavelength
$\lambda = \frac{2\pi}{k_F}$ and the characteristic time scale $\tau_0 = \frac{d }{v_F}$. The quantity $\tau_0$ represents the time required for a free electron (without quantum effects) to traverse the barrier.	

\section{Numerical Analysis}\label{NAA}

\begin{figure}[ht]
	\centering
	\subfloat[$V_0=100$~meV, $d=40$~nm]{
		\centering
		\includegraphics[scale=0.27]{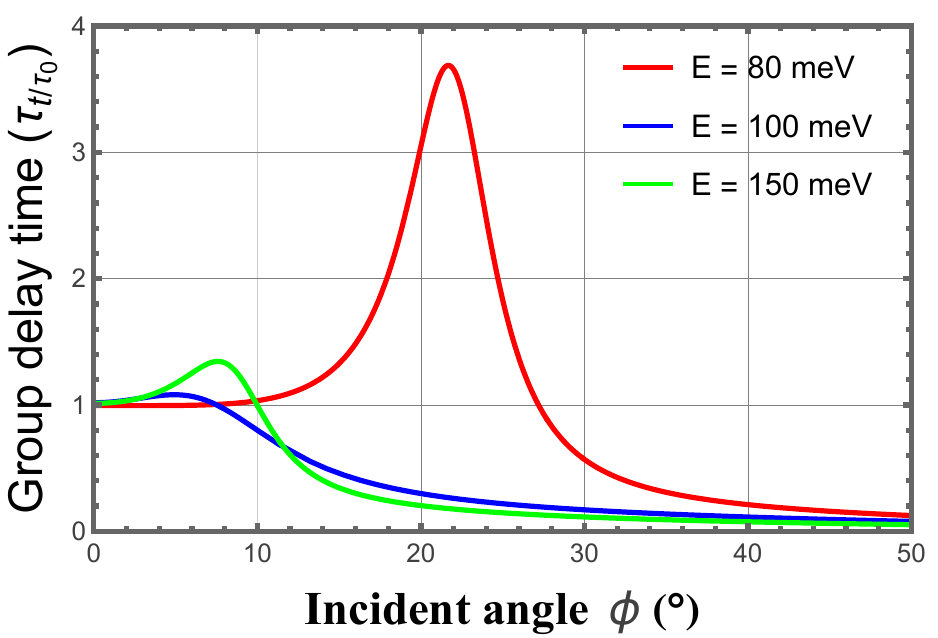}}
	\subfloat[$E=120$~meV, $V_0=100$~meV]{
		\centering\includegraphics[scale=0.27]{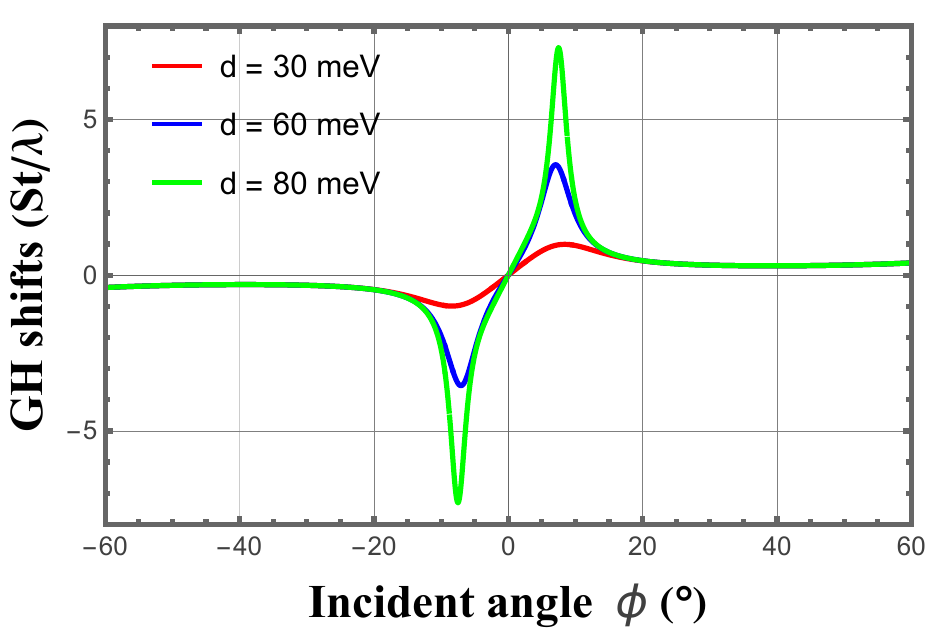}
	}  \\
	\subfloat[$V_0=120$~meV, $d=40$~nm]{
		\centering\includegraphics[scale=0.27]{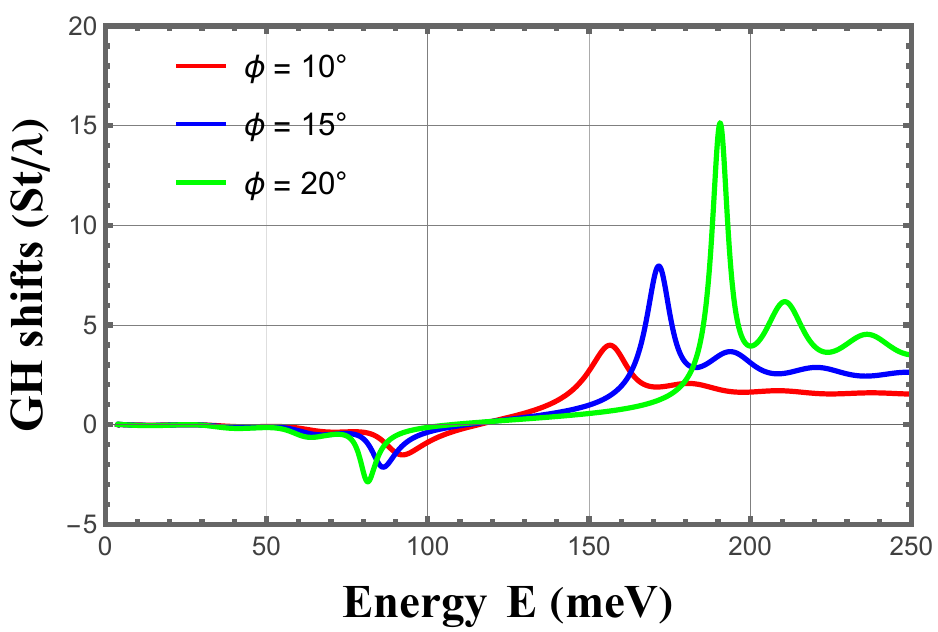}
	} 
	\subfloat[$E=150$~meV, $d=40$~nm]{
		\centering\includegraphics[scale=0.27]{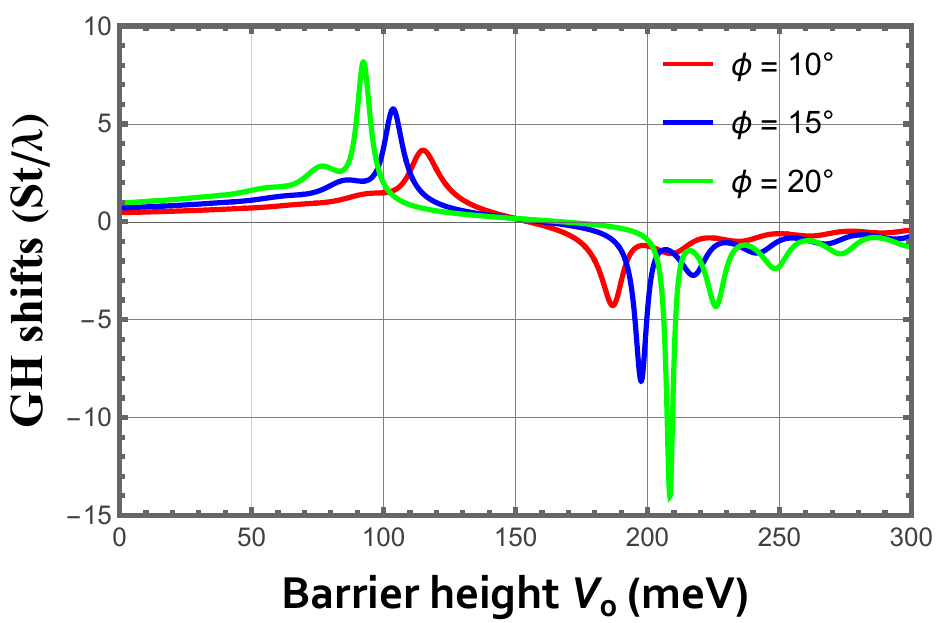}
	} 
	\caption{The GH shifts in transmissions $S_t/\lambda$ versus the incident angle  $\phi$  (a,b), energy $E$ (c), and barrier height $V_0$
		(d).}
\label{fig3}
\end{figure}

In Fig.~\ref{fig3} we illustrates the influence of the incident angle $\phi$, the barrier width $d$, incident energy $E$, and the  barrier height $V_0$ on the GH shifts in transmission. In  Fig. \ref{fig3}{{(a,b)}}, the GH shifts $S_t$ is
shown versus the incident angle $\phi$ by selecting different energy values  $E=(120, 150, 180)$ meV,  {$V_0=100$} meV, and  $d=80$ nm in Fig.~\ref{fig3}({{a}}), whereas  different
width values  $d=(30, 60, 80)$~nm, 
$V_0=100$ meV, and $E=120 $ meV in Fig.~\ref{fig3}({b}). In both figures, the sign change of $S_t$ at normal incidence $(\phi = 0)$ originates from the transverse symmetry of the silicene
barrier system. At this point, the transverse wave vector vanishes
($k_y = 0$), leading to zero lateral displacement of the transmitted
electron beam, and therefore $S_t$ is exactly zero.
For oblique incidence $(\phi \neq 0)$, the transverse momentum $k_y$ acquires opposite signs for positive and negative incident angles. 
This asymmetry gives rise to positive and negative $S_t$ on either side of $\phi = 0$, satisfying the relation 
$S_t(+\phi) = -S_t(-\phi)$. This behavior reflects the Dirac-like nature of charge carriers in
silicene and is closely connected to the transition between Klein
tunneling and classical transport regimes. In particular, negative $S_t$ corresponds to the Klein tunneling regime, where the group velocity is antiparallel to the wave vector, whereas positive $S_t$ arise in the classical regime, where the group velocity and wave vector are parallel. 
The presence of the energy gap in silicene primarily influences the magnitude and
resonant behavior of $S_t$, without altering the fundamental
symmetry-induced sign change at normal incidence. 
As shown in Fig.~\ref{fig3}(a), increasing the incident energy modifies both the amplitude and the angular positions of the $S_t$ resonance peaks, owing to the energy-dependent phase accumulation within the barrier. In contrast, Fig.~\ref{fig3}(b) demonstrates that increasing the barrier width enhances the magnitude of $S_t$ without shifting the resonance positions, indicating that the barrier width acts as a scaling factor for the accumulated phase rather than altering the resonance conditions.
In Fig.~\ref{fig3}(c), we plot $S_t$ as a function of the incident energy $E$ for different incident angle values  $\phi = (10^\circ, 15^\circ, 20^\circ)$, $V_0 = 120~\mathrm{meV}$, and $d = 40~\mathrm{nm}$. We observe  that $S_t$ changes sign in the vicinity of
the Dirac point $E = V_0$ and becomes significant at certain resonance
points. Indeed, this sign change of $S_t$ originates from the
fact that the Dirac point $E = V_0$ corresponds to the transition between
the Klein tunneling regime ($E < V_0$) and the classical motion regime
($E > V_0$). It can also be seen that the absolute value of $S_t$ increases as the incidence angle $\phi$ increases. 
In addition, the control of $S_t$ through the applied electrostatic potential is displayed in  Fig.~\ref{fig3}(d),
where $d = 80\ \text{nm}$, $E = 150\ \text{meV}$, and $\phi = (10^\circ$, $15^\circ$, $20^\circ$),  showing that $S_t$ can be controlled by changing $V_0$. The point $V_0 = E$ indicates a singularity in the transmission spectrum, which represents the zero modes for Dirac operators and leads to the emergence of new Dirac points.
Such a point separates the two regions of positive and negative refraction.
In the cases $V_0 < E$ and $V_0 > E$, $S_t$ is respectively in the forward and backward directions,
due to the fact that the signs of the group velocity are opposite.
It is therefore suggested that the incident energy can be selectively determined through these tunable beam shifts.

\begin{figure}[ht]
\centering
\subfloat[$V_0=120$~meV, $d=40$~nm]{
	\centering
	\includegraphics[scale=0.27]{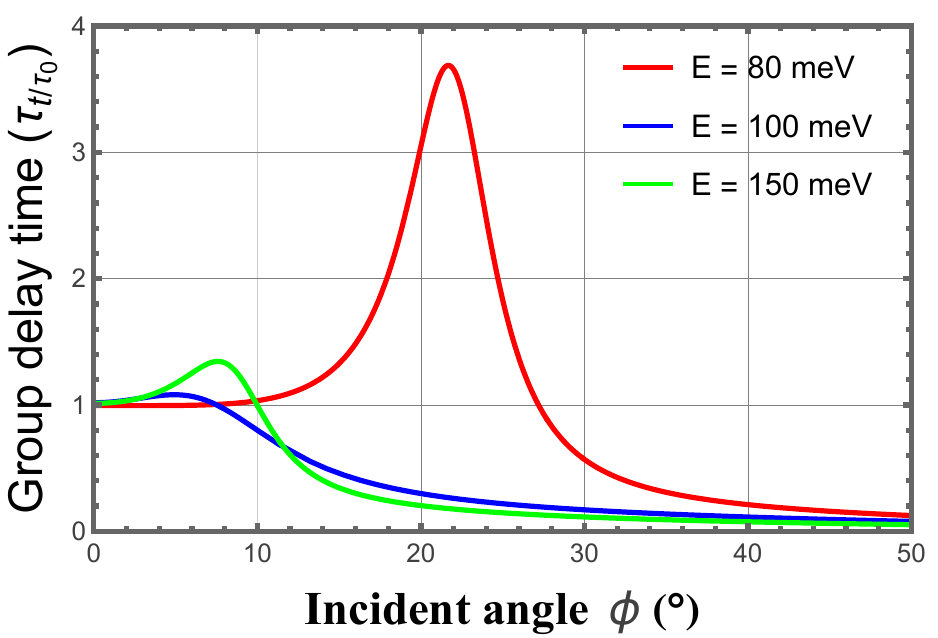}}
\subfloat[$V_0=120$~meV, $d=60$~nm]{
	\centering\includegraphics[scale=0.27]{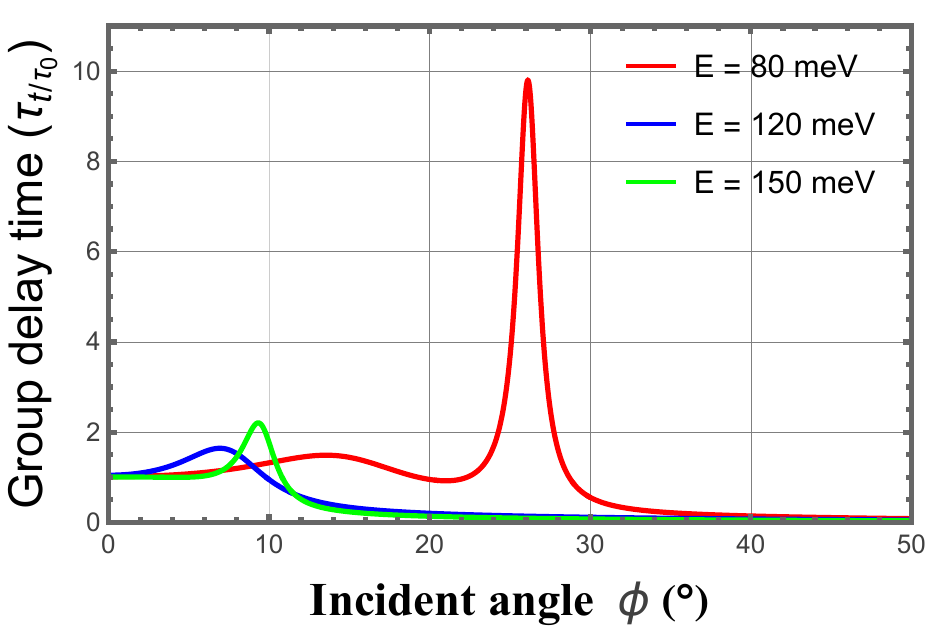}
}  \\
\subfloat[$V_0=120$~meV, $d=80$~nm]{
	\centering\includegraphics[scale=0.27]{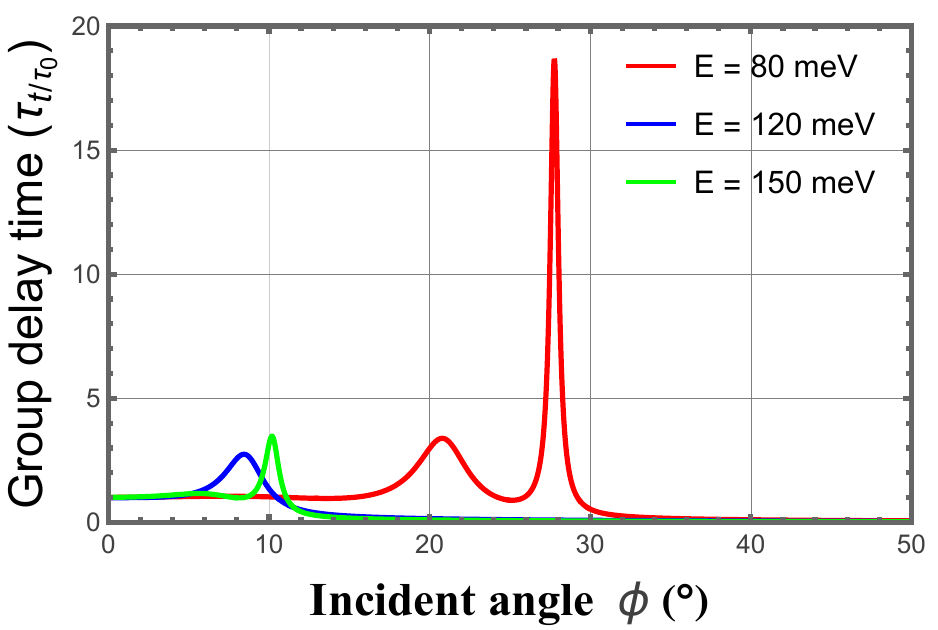}
} 
\subfloat[$V_0=120$~meV, $d=100$~nm]{
	\centering\includegraphics[scale=0.27]{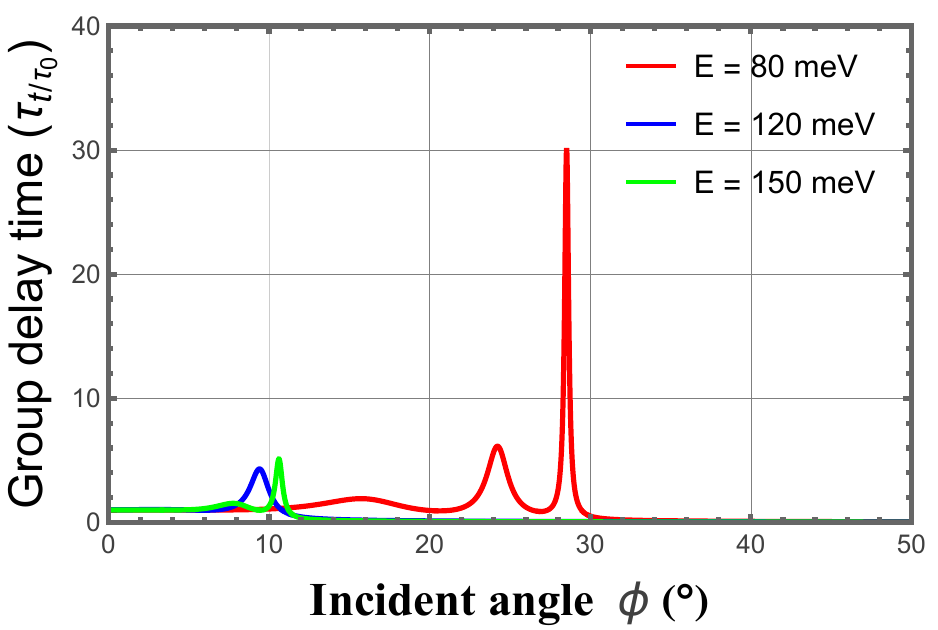}
} 
\caption{The group delay time in transmission $\tau_{t}/\tau_{0}$ versus the
	incident angle $\phi$.}
\label{fig4}
\end{figure}

Figure~\ref{fig4} shows the group delay time in transmission $\tau_t/\tau_0$ as a function of the incident angle $\phi$ for different physical parameters. The barrier height is fixed at $V_0 = 100~\text{meV}$, where the incident energy takes the values $E = ({80,100,150})\ \text{meV}$, and four different barrier widths $d = (40,60,80,100)~\text{nm}$ are considered, corresponding to Figs.~\ref{fig4}(a,b,c,d), respectively. 
It is observed that at normal incidence ($\phi = 0$), the particles traverse the barrier with the Fermi velocity $v_F$, which corresponds to $\tau_t/\tau_0 = 1$. However, as  $\phi$ increases,  $\tau_t/\tau_0$ increases progressively until it reaches a maximum, after which it decays exponentially and eventually tends to zero at $\phi = 40^\circ$. This behavior can be explained by the fact that the wave vector inside the barrier becomes imaginary, leading to an evanescent wave function in the barrier region. Moreover, it was found that $\tau_t/\tau_0$ increases with the increase of the incident energy $E$. In contrast, increasing the barrier width $d$ makes the resonance peaks narrower while increases their number. This reflects the enhanced  effects of  quantum interference and the formation of additional quasi-bound states inside the barrier.

\begin{figure}[ht]
\subfloat[$\phi=10^{\circ}$, $d=40$~nm]{
\centering
\includegraphics[scale=0.27]{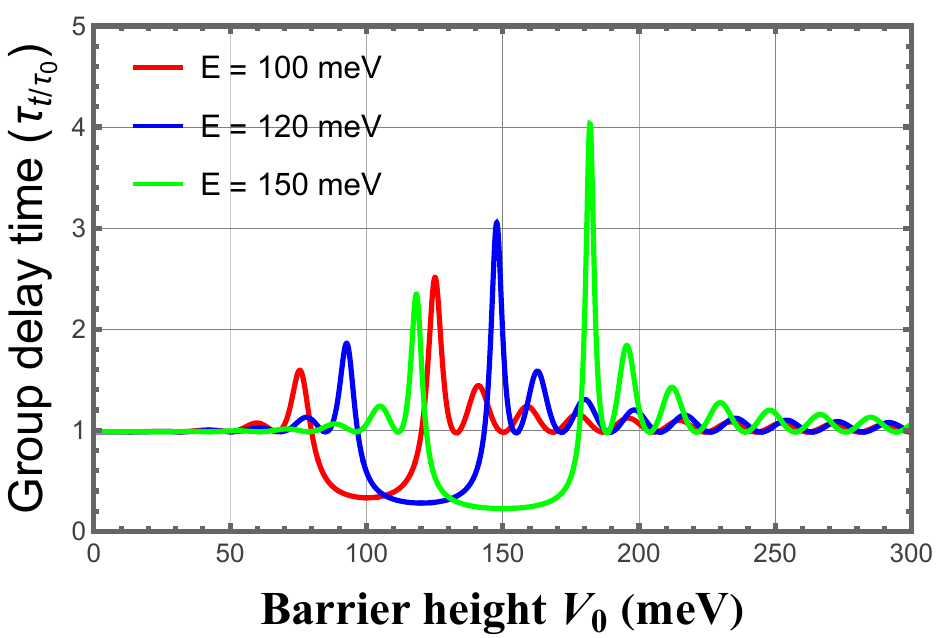}}
\subfloat[$\phi=15^{\circ}$, $d=40$~nm]{
\centering\includegraphics[scale=0.27]{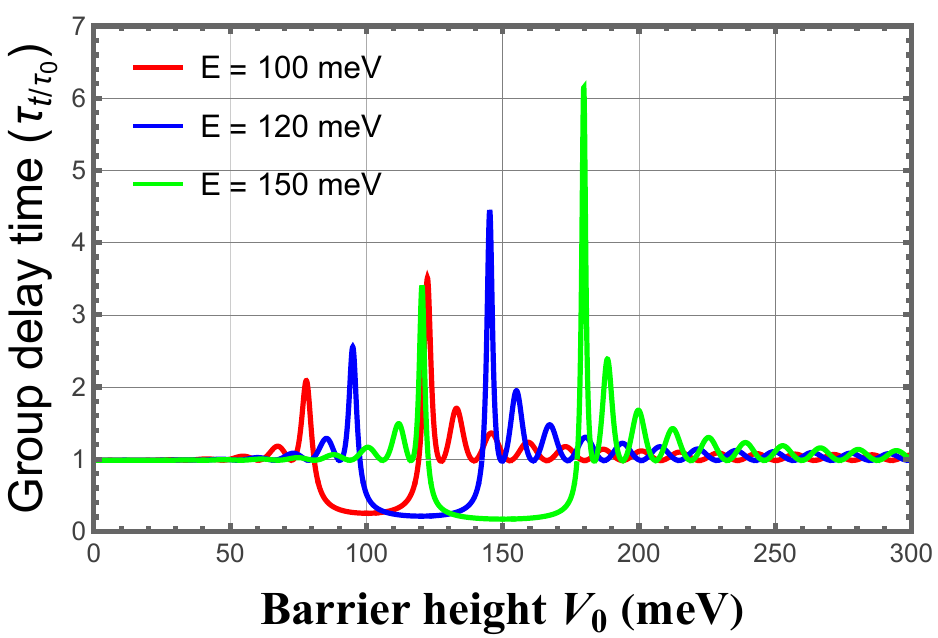}
}  \\
\subfloat[$\phi=20^{\circ}$, $d=40$~nm]{
\centering\includegraphics[scale=0.27]{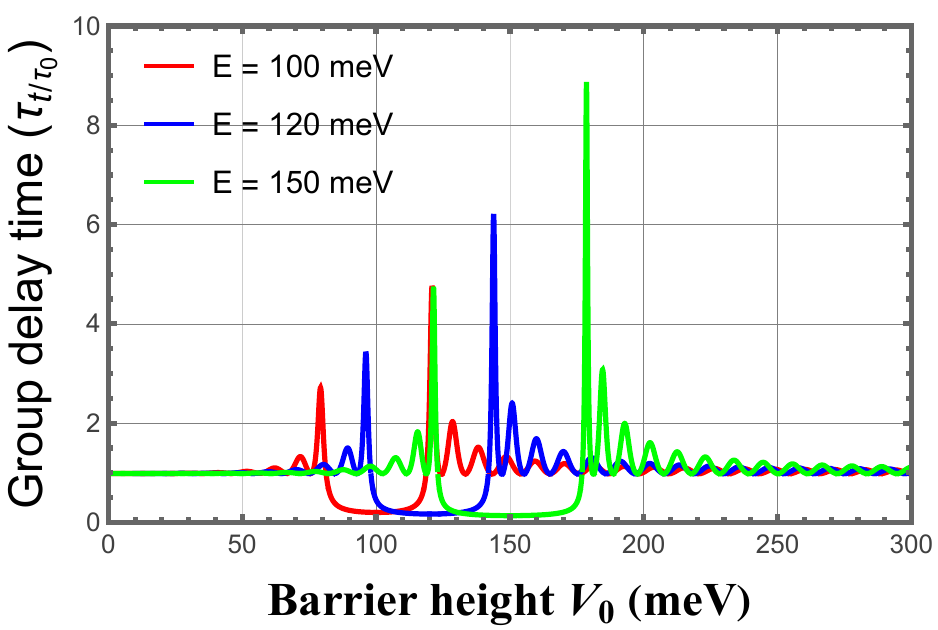}
} 
\subfloat[$\phi=30^{\circ}$, $d=40$~nm]{
\centering\includegraphics[scale=0.27]{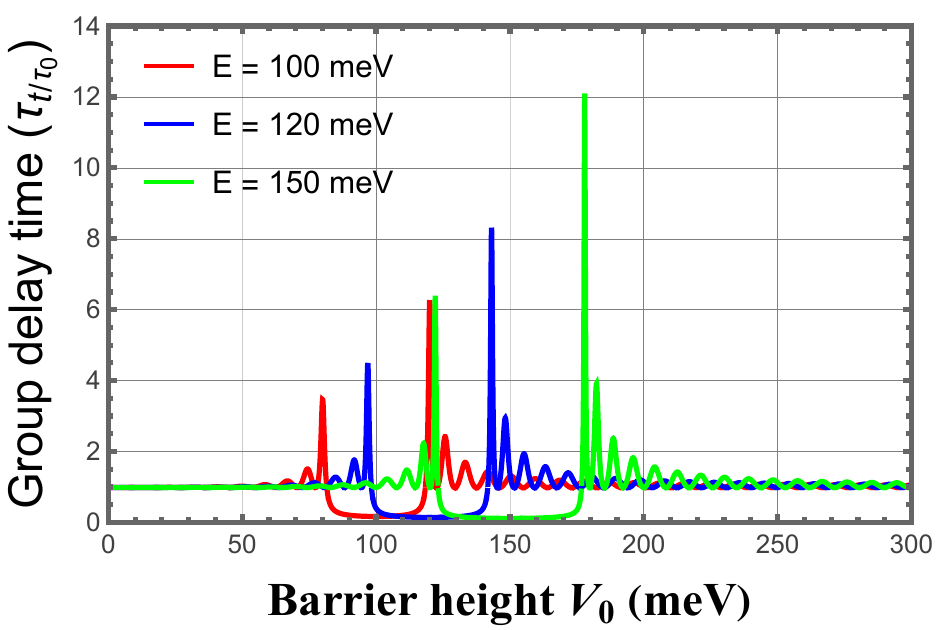}
} 
\caption{The group delay
time in transmission $\tau_{t}/\tau_{0}$ versus the barrier height $V_0$ for barrier width $d=40$~nm, different values of energy $E$ and incident angle $\phi$.}
\label{fig5}
\end{figure}

In  Fig.~\ref{fig5}, we show the group delay time
in transmissions $\tau_t/\tau_0$ as a function of the barrier height $V$ for different physical parameters. The barrier width is fixed at $d=40$ nm, the incident energy is chosen as 
$E={80,100,150}$ meV and four different incident angles, $\phi = {10^\circ, 15^\circ, 20^\circ, 30^\circ}$, are shown in Figs.~\ref{fig5}{(a,b,c,d)}, respectively. For low height values, the carriers propagate through the barrier with the Fermi velocity, yielding $\tau_t/\tau_0 = 1$. When $V_0 < E$,  $\tau_t/\tau_0$ initially increases and then decreases, reaching a minimum at $V_0 = E$, which corresponds to the Dirac point. This minimum originates from the transition between the propagative and tunneling regimes. Under this condition, the density of states inside the barrier is strongly modified, leading to a suppression of the group delay time.
In the regime $E < V_0$, $\tau_t/\tau_0$ increases again and exhibits a pronounced oscillatory behavior. These oscillations arise from quantum interference effects inside the barrier, which can be interpreted as Fabry-P\'erot–type resonances due to multiple reflections of Dirac fermions at the barrier interfaces. The constructive and destructive interferences give rise to resonant transmission peaks, corresponding to the formation of quasi-bound states within the barrier region.
Regarding the effect of the incident energy and angle, the group delay time increases with increasing energy $E$ and incident angle $\phi$. In particular, larger incident angles enhance the confinement of carriers inside the barrier, leading to higher and sharper resonance peaks. This behavior reflects a stronger phase accumulation and a more efficient Fabry-P\'erot interference mechanism. Overall, the interplay between the barrier parameters and the position of the Dirac point allows for an efficient control of the temporal dynamics of Dirac fermions in silicene.

\begin{figure}[ht]
\centering
\subfloat[$\phi=10^{\circ}$, $d=60$~nm]{
\centering
\includegraphics[scale=0.27]{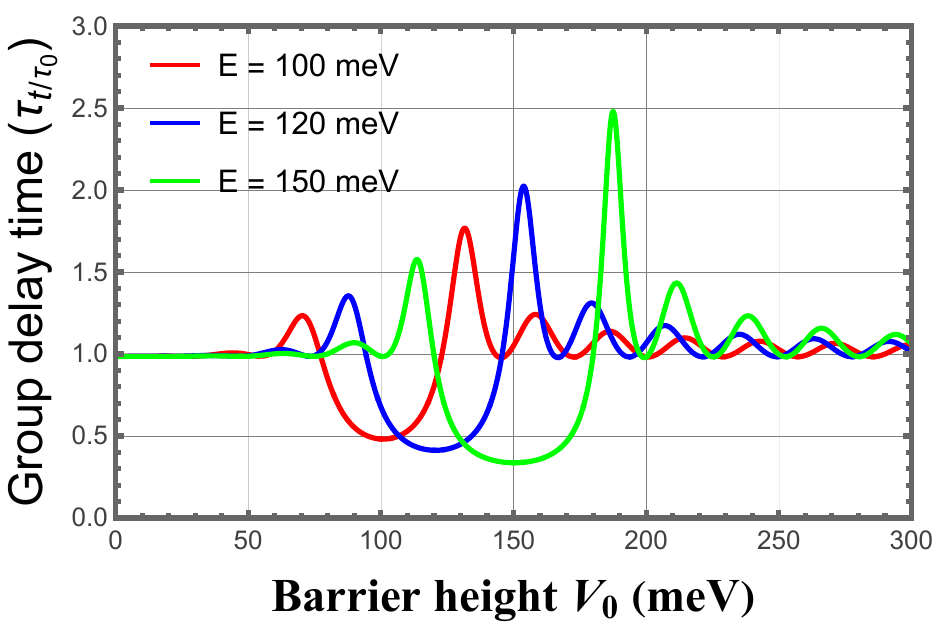}}
\subfloat[$\phi=10^{\circ}$, $d=80$~nm]{
\centering\includegraphics[scale=0.27]{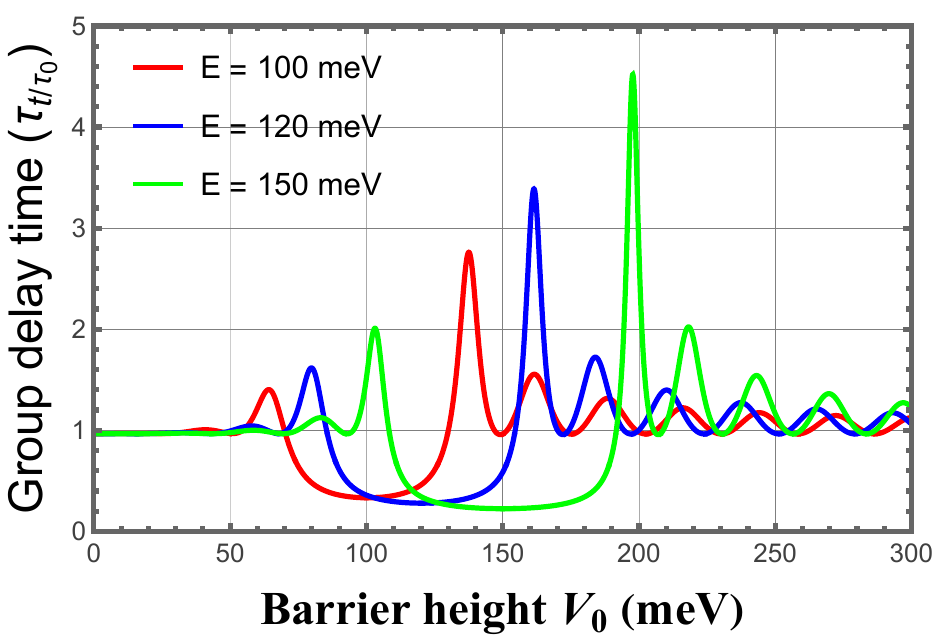}
}  \\
\subfloat[$\phi=10^{\circ}$, $d=100$~nm]{
\centering\includegraphics[scale=0.27]{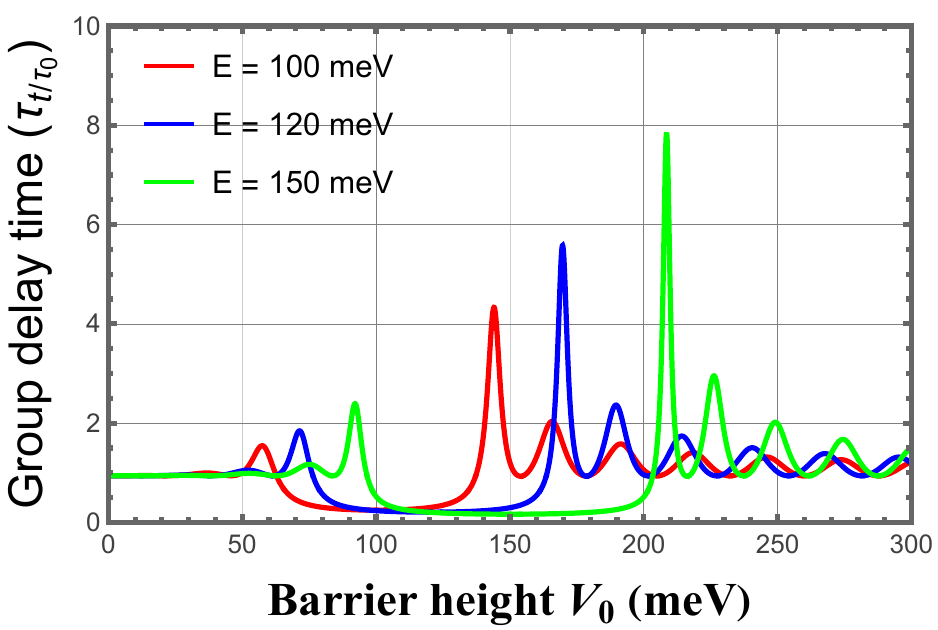}
} 
\subfloat[$\phi=10^{\circ}$, $d=120$~nm]{
\centering\includegraphics[scale=0.27]{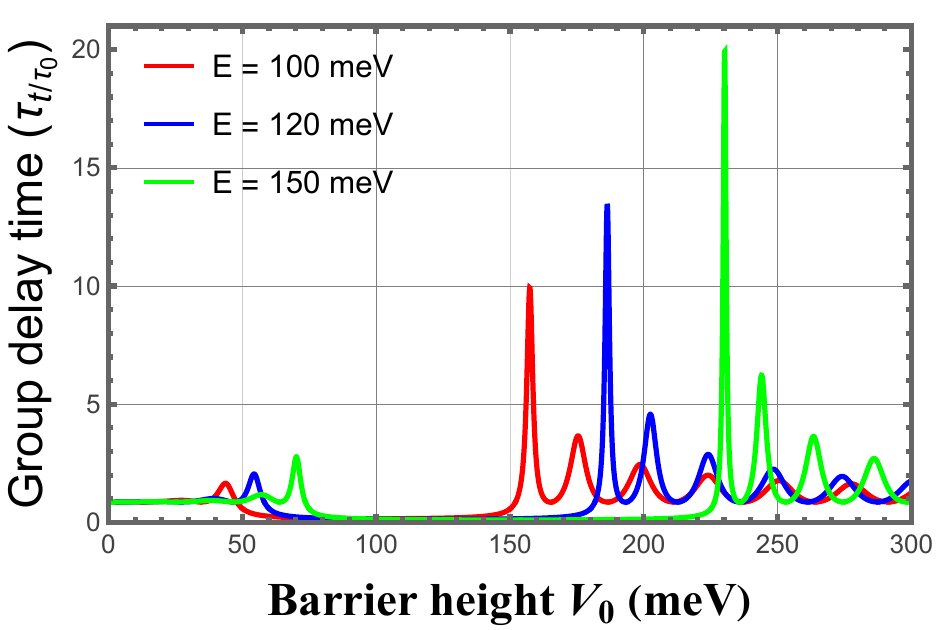}
} 
\caption{The group delay
time in transmission $\tau_{t}/\tau_{0}$ versus the barrier height $V_0$ for incident angle $\phi=10^\circ$,   different values
of energy $E$ and barrier width $d$.}
\label{fig6}
\end{figure}

Figure~\ref{fig6} shows that varying the incident energy $E$ and the barrier width $d$, while keeping the incident angle at $\phi=10^\circ$ yields behavior qualitatively similar to that discussed Fig.~\ref{fig5}. 
In particular, the group delay in transmission $\tau_t/\tau_0$ exhibits oscillatory features associated with transmission resonances in the propagation regime. 
Its magnitude is strongly influenced by the barrier parameters. The main difference lies in how the resonance structure is modulated. Increasing $E$ or $d$ alters the position and amplitude of the peaks, reflecting changes in the interference conditions inside the barrier. These results demonstrate that the group delay time in silicene-based barriers can be efficiently tuned by the barrier height, incident energy, and barrier width.

\begin{figure}[ht]
\centering
\subfloat[$\phi=10^{\circ}$, $d=40$~nm]{
\centering
\includegraphics[scale=0.27]{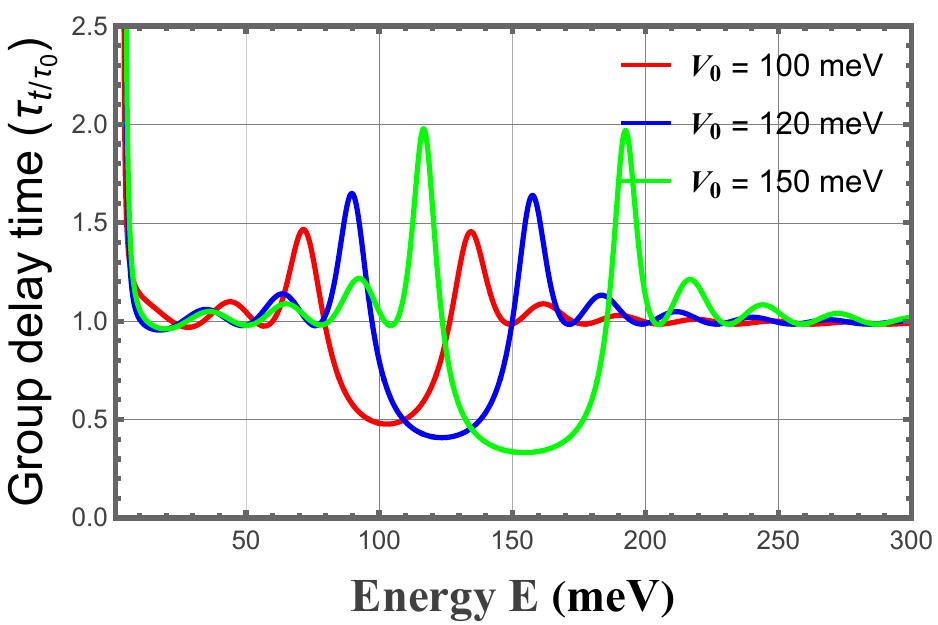}}
\subfloat[$\phi=15^{\circ}$, $d=40$~nm]{
\centering\includegraphics[scale=0.27]{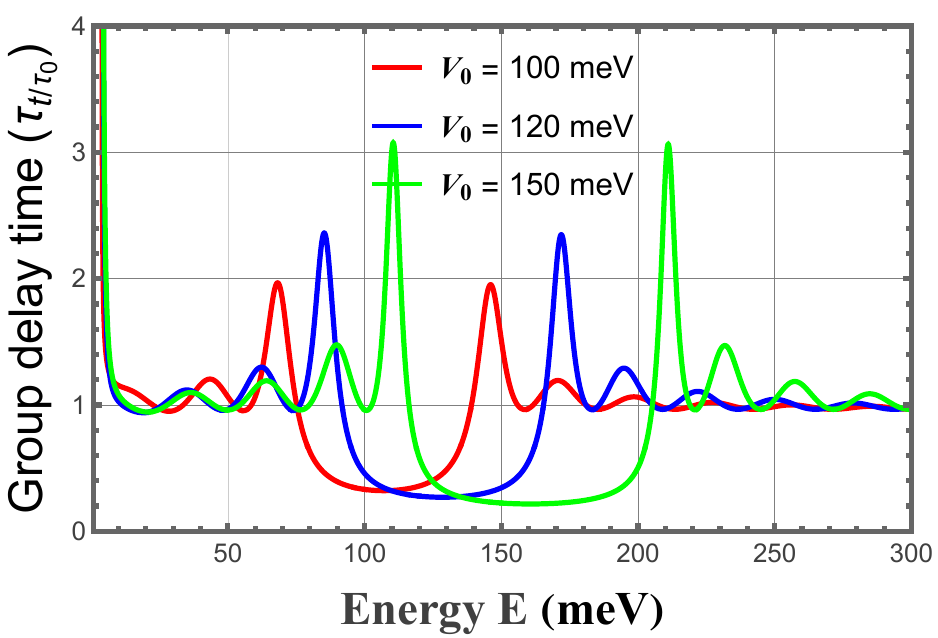}
}  \\
\subfloat[$\phi=20^{\circ}$, $d=40$~nm]{
\centering\includegraphics[scale=0.27]{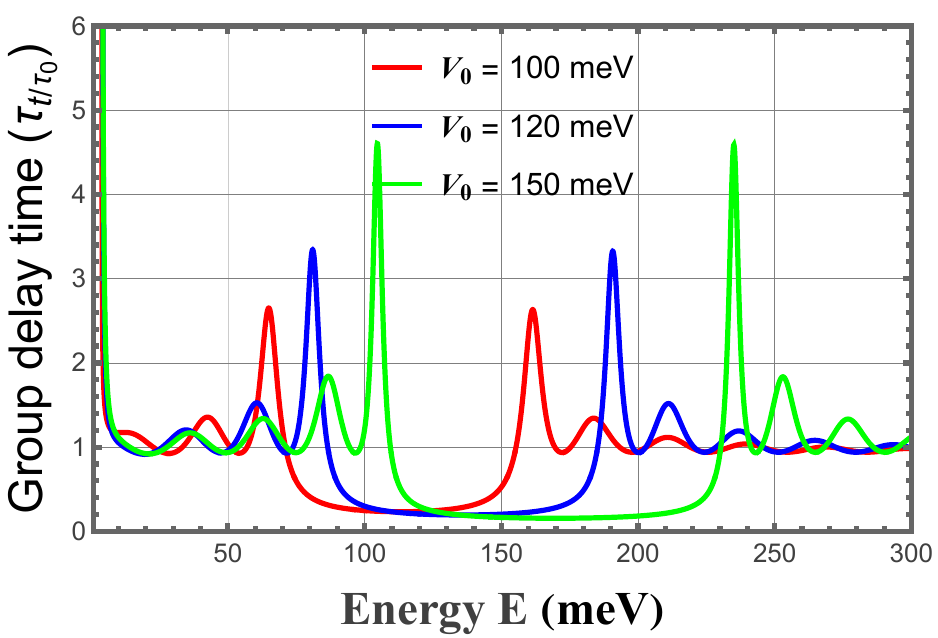}
} 
\subfloat[$\phi=30^{\circ}$, $d=40$~nm]{
\centering\includegraphics[scale=0.27]{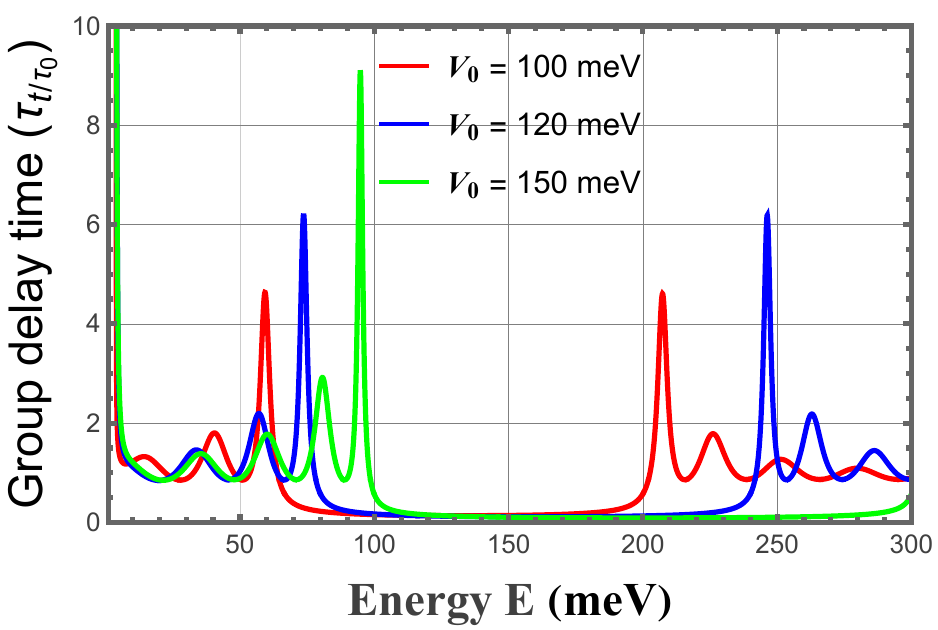}
} 
\caption{The group delay
time in transmission $\tau_{t}/\tau_{0}$ versus the incident energy $E$ for barrier width $d=40$~nm, different values
of the barrier height $V_0$ and incident angle $\phi$.}
\label{fig7}
\end{figure}

We show in Fig.~\ref{fig7} the group delay time in transmission  $\tau_t/\tau_0$ versus the incident energy $E$ for three values of barrier heights $V_0= (100, 120, 150)$ meV and four incident angles $\phi= (10^\circ, 15^\circ, 20^\circ, 30^\circ$), while keeping the barrier width fixed at $d=40$~nm. At low energies, the quasiparticles traverse the barrier with the Fermi velocity, resulting in $\tau_t/\tau_0 \approx 1$. As both $E$ and $\phi$ increase, the oscillatory pattern becomes more pronounced, indicating stronger transmission resonances. Furthermore, the amplitude of the group delay time increases with the barrier height $V_0$, showing that higher potential barrier prolongs the group delay time of the quasiparticles inside the barrier. A vanishing group delay time corresponds to total reflection, whereas pronounced peaks are associated with resonant tunneling. The results also emphasize that the Dirac point $E=V_0$ can be shifted by the electrostatic barrier.

\begin{figure}[ht]
\centering
\subfloat[$\phi=10^{\circ}$, $d=60$~nm]{
\centering
\includegraphics[scale=0.27]{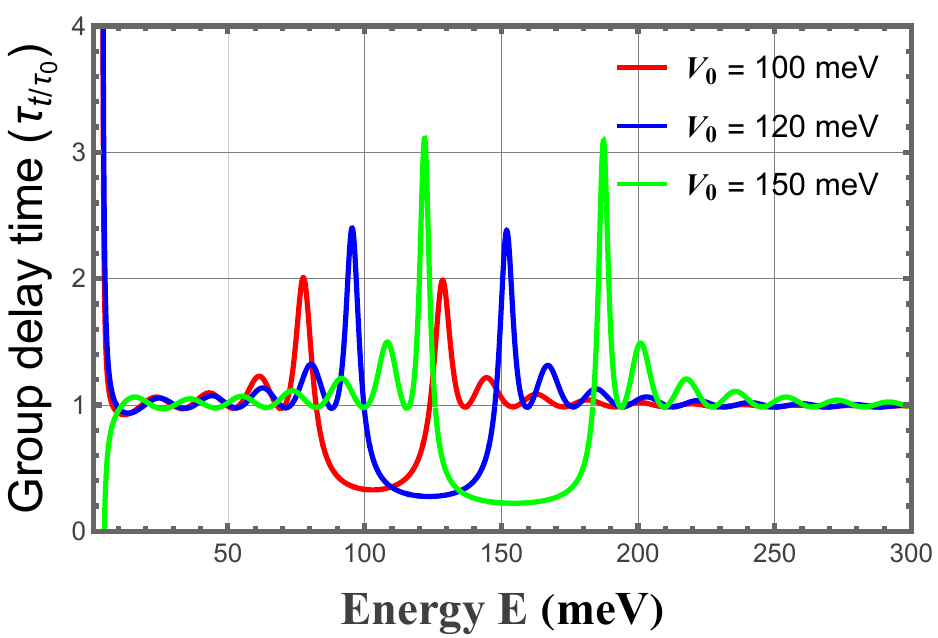}}
\subfloat[$\phi=10^{\circ}$, $d=80$~nm]{
\centering\includegraphics[scale=0.27]{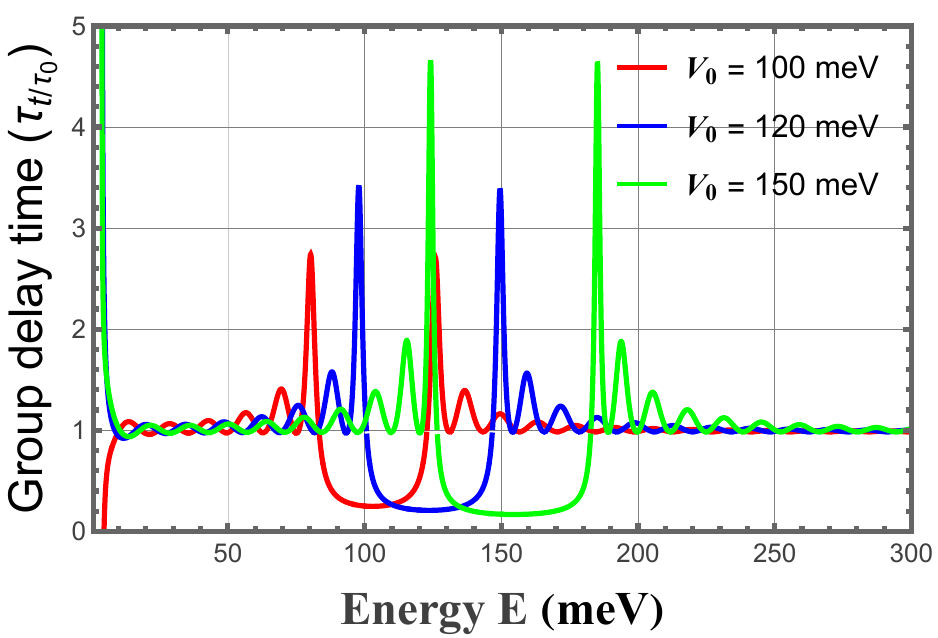}
}  \\
\subfloat[$\phi=10^{\circ}$, $d=100$~nm]{
\centering\includegraphics[scale=0.27]{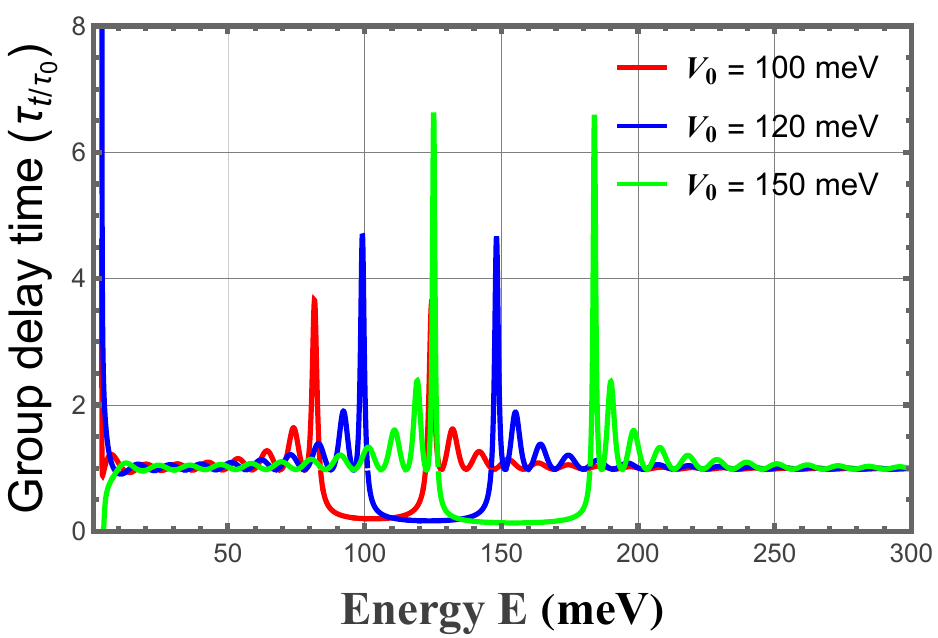}
} 
\subfloat[$\phi=10^{\circ}$, $d=120$~nm]{
\centering\includegraphics[scale=0.27]{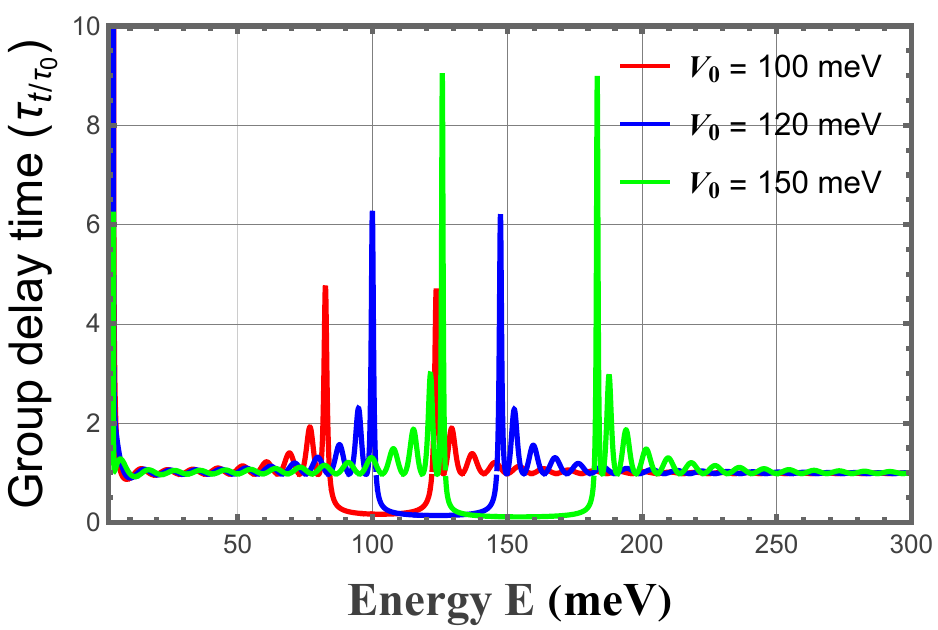}
} 
\caption{The group delay
time in transmission $\tau_{t}/\tau_{0}$ versus the incident energy $E$  for incident angle $\phi=10^\circ$,  different values
the barrier height $V_0$ and barrier width $d$. }
\label{fig8}
\end{figure}

In Fig.~\ref{fig8}, the same general trends observed in Fig.~\ref{fig7} are also observed when varying the barrier height $V_0$ and width $d$, while keeping the incident angle fixed. In particular, the group delay time exhibits pronounced oscillations due to transmission resonances, and the overall behavior remains consistent with the previous discussion. However, increasing  $V_0$ or  $d$ leads to a noticeable enhancement of the group delay time, reflecting a longer group delay time of quasi-particles within the silicene barrier. These results confirm that the electrostatic barrier can significantly modulate the group delay time and shifts the position of the Dirac point $E=V_0$. 

\begin{figure}[ht]
\centering
\subfloat[$\phi=15^{\circ}$, $V_0=120$~meV]{
\centering
\includegraphics[scale=0.27]{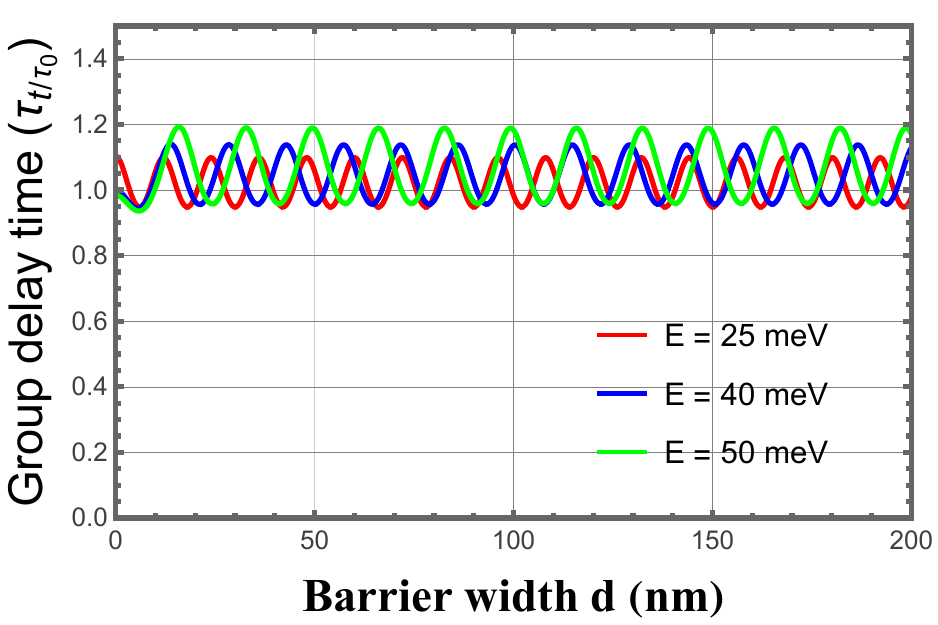}}
\subfloat[$\phi=25^{\circ}$, $V_0=120$~meV]{
\centering\includegraphics[scale=0.27]{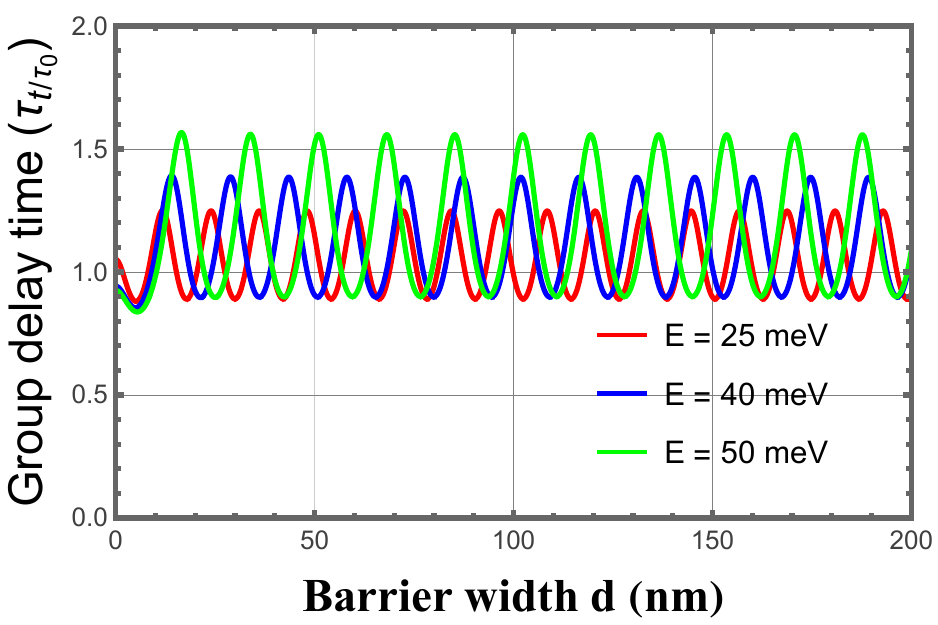}
}  \\
\subfloat[$\phi=30^{\circ}$, $V_0=120$~meV]{
\centering\includegraphics[scale=0.27]{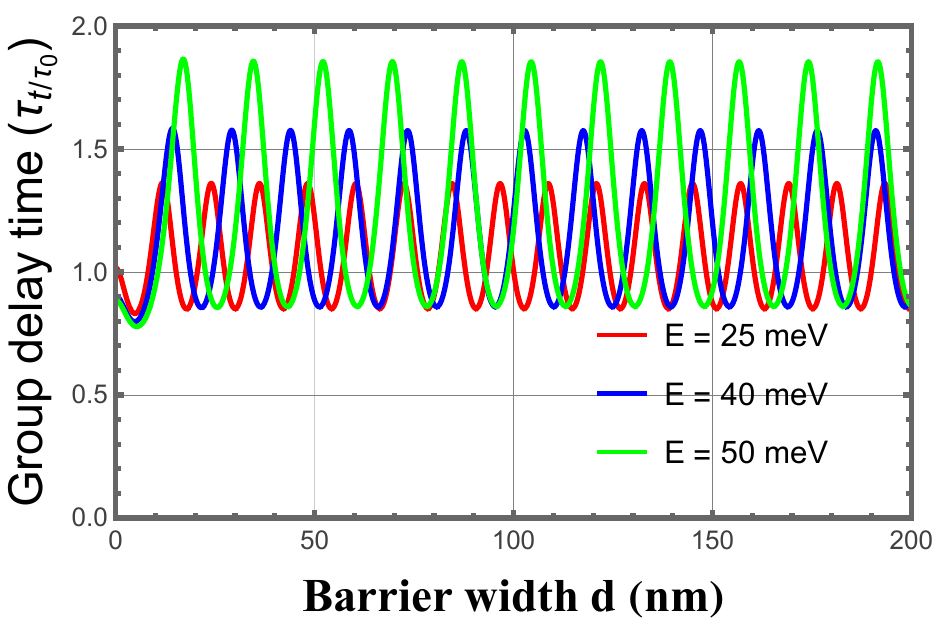}
} 
\subfloat[$\phi=40^{\circ}$, $V_0=120$~meV]{
\centering\includegraphics[scale=0.27]{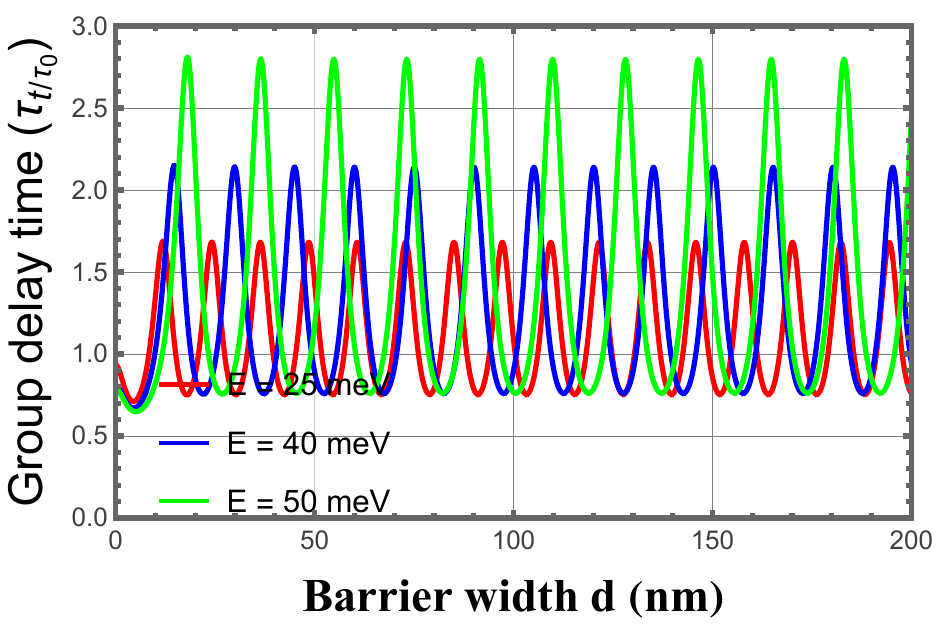}
} 
\caption{The group delay
time in transmission $\tau_{t}/\tau_{0}$ versus the barrier width $d$  for the barrier height $V_0=120 $ meV, different values of incident angle $\phi$
and energy $E$.}
\label{fig9}
\end{figure}

Figure~\ref{fig9} presents the group delay time $\tau_t/\tau_0$ versus the barrier width $d$, while keeping the barrier height fixed at $V_0=120~\text{meV}$. The analysis is performed for three incident energies $E = (100, 120,150)~\text{meV}$ and incident angles $\phi = (10^\circ,\,15^\circ,\,20^\circ,\,30^\circ)$.
Our results reveal that $\tau_t/\tau_0$ increases with the increase of $d$, which suggest that the particles spend longer time inside the barrier region due to multiple internal reflections. In addition, $\tau_t/\tau_0$ increases with incident energy $E$ and incident angle $\phi$. The higher incident energy increases the number of resonant modes within the barrier, whereas the larger incident angle increases the effective path length and modifies the resonance conditions.
Furthermore, both the number and amplitude of resonance peaks increase with $d$, $E$, and $\phi$. These resonances arise from constructive interference of the wave function within the barrier, corresponding to quasi-bound states that significantly enhance the group delay time. Consequently, a rectangular potential barrier in silicene offers an efficient mechanism for controlling the group delay time through quantum interference effects.

\hspace{0.1mm}

\section{Silicene vs graphene}\label{SSGG}

The present work builds upon our earlier investigations of Goos–Hänchen (GH) shifts and group delay time in graphene under various external constraints \cite{Jel1,Jel2,Jel3}, and extends our more recent study of tunneling phenomena of Dirac fermions in silicene through electrostatic potential barriers \cite{JellalAP2024}. This progression naturally allows for a direct comparison between graphene and silicene, two prototypical Dirac materials, with particular emphasis on how their intrinsic properties influence lateral beam shifts and temporal delay effects during barrier scattering.

In graphene, GH shifts and group delay times arise primarily from phase variations of the transmission and reflection amplitudes at the barrier interfaces. Graphene exhibits gapless band structure together with its very low intrinsic spin–orbit coupling, which leads to spin-degenerate properties that depend mainly on the geometric parameters of incident angle and barrier width and barrier height. Our work~\cite{Jel1,Jel2,Jel3} shows that external laboratory forces such as electrostatic gating and strain and magnetic fields can be used to control both GH shifts and group delay time. The resulting effects show strong dependency on valley characteristics which do not provide complete spin resolution while the absence of an intrinsic band gap restricts the ability to control resonant states within the barrier region.

Silicene, by contrast, exhibits a qualitatively different behavior due to its buckled honeycomb lattice and relatively strong intrinsic spin–orbit coupling. As shown in~\cite{JellalAP2024}, the presence of a finite and electrically tunable band gap strongly modifies the phase accumulation of Dirac fermions inside the barrier. This, in turn, has a pronounced impact on both GH shifts and group delay times. In particular, the intrinsic spin–orbit interaction naturally lifts the spin degeneracy, leading to spin-dependent lateral shifts and time delays even in the absence of external magnetic fields. As a result, silicene displays richer and more complex GH shifts and group-delay patterns compared to graphene.

Another important distinction lies in how quasi-bound states form within the barrier. Such states are generally weak and depend heavily on external constraints in graphene, but they appear as more stable states in silicene because the material has a finite gap and stronger barrier confinement. These quasi-bound states create sharp resonances in group delay time and produce large GH shift oscillations, which can be efficiently controlled by adjusting the barrier parameters or applying an external electric field. This ability to control wave-packet dynamics represents a significant advantage of silicene over pristine graphene, which lacks this feature.

Graphene is a clean and adaptable research platform that allows scientists to study fundamental aspects of graphene (GH) shifts and group delay phenomena. However, silicene provides researchers with more efficient methods to control these phenomena due to its material characteristics. A comparison of the two materials shows that silicene expands the physical properties of graphene and creates new pathways for research into spin- and valley-dependent transport. Silicene-based heterostructures demonstrate the ability to support advanced spintronic and valleytronic technologies, which require precise control of charge-carrier movement in space and time.

{In addition to electrostatic barriers, transport properties of graphene have been shown to be affected by an external dressing field, such as strong electromagnetic radiation or periodic driving~\cite{Kibis2010, Anwar2020, Iurov2013}. These fields renormalize the electronic spectrum, potentially inducing effective band gaps, modifying Klein tunneling conditions, and altering the phase of transmission amplitudes. As a result, both Goos–Hänchen shifts and group delay times can be tuned via light–matter interactions. These findings highlight the richness of wave-packet dynamics in Dirac materials and complement the electrostatic control mechanisms investigated in the present silicene system.}

\section{Conclusion}\label{CCC}

{We have presented a comprehensive study of the Goos–Hänchen (GH) shifts and the group delay time of Dirac fermions traversing a rectangular electrostatic potential barrier in silicene. By systematically analyzing their dependence on the incident angle, barrier height, barrier width, and incident energy, we have clarified how quantum interference and barrier-induced confinement influence both the spatial and temporal propagation of charge carriers in this two-dimensional Dirac material.
Our results show that the GH shifts exhibit pronounced oscillatory behavior arising from multiple reflections and quantum interference within the barrier. Both the amplitude and number of oscillations increase with incident energy, barrier width, and incidence angle, reflecting enhanced lateral displacement as phase accumulation inside the barrier becomes stronger. These features are closely linked to the formation of resonant and quasi-bound states, which selectively enhance the transverse displacement of the transmitted wave packet.


The group delay time shows a complementary behavior, with significant resonant features related to the same quasi-bound states responsible for the Goos–Hänchen oscillations. In our calculations, we observe that the delay time increases with barrier width, incident energy, and incidence angle, reflecting the longer dwell time of carriers inside the barrier. Conversely, higher barrier heights reduce the delay because the transmission probability of carriers decreases. These results highlight the connection between spatial and temporal aspects of Dirac fermion transport, as both arise from phase accumulation and interference within the electrostatic barrier.
Furthermore, the analogy with the GH effect suggests that electrostatic barriers can be used to simultaneously control carrier displacement and delay. This is an important feature for applications in electron optics and Dirac-based devices, where controlling both the position and timing of carriers is critical. These findings also indicate that potential barriers provide a versatile platform for exploring transport features in two-dimensional Dirac systems such as silicene.


Overall, our study provides a clear and coherent physical picture of how electrostatic barriers can be used to manipulate both the lateral and temporal dynamics of Dirac fermions in silicene. The ability to tune the Goos–Hänchen shifts and group delay through external parameters opens promising opportunities for controlling electron-beam steering, signal timing, and interference effects at the nanoscale. These results underscore the potential of silicene-based structures for tunable nanoelectronic and spintronic devices, as well as other two-dimensional Dirac materials, where precise control over carrier propagation is essential for device functionality.}

\end{document}